\documentclass[prl,a4paper]{revtex4}

\usepackage{amsmath,amssymb}
\usepackage{epsfig}

\begin{document}
\title{Decisive test for the Pomeron at Tevatron}

\author{A. Kup\v{c}o\footnote{On leave from Institute of Physics, Center for
Particle Physics,
Prague, Czech Republic}
}
\affiliation{DSM/DAPNIA/SPP, CEA Saclay,
  91191 Gif-sur-Yvette Cedex, France.}
\email{kupco@fzu.cz, royon@hep.saclay.cea.fr} 
\author{R. Peschanski}
\affiliation{Service de physique th\'eorique\footnote{URA 2306, unit\'e de 
recherche associée au CNRS.}, CEA/Saclay,
  91191 Gif-sur-Yvette Cedex, France}
\email{pesch@spht.saclay.cea.fr}
\author{C. Royon}
\affiliation{DSM/DAPNIA/SPP, CEA Saclay,
  91191 Gif-sur-Yvette Cedex, France.}

\begin{abstract}
We propose a new measurement to be performed at the Tevatron which can be
decisive to distinguish between Pomeron-based and soft color interaction
models of hard diffractive scattering.
\end{abstract}

\maketitle

{\bf 1. Introduction} 
The hard diffraction phenomena revealed at HERA \cite{h1pom} has put a 
new light 
on the longlasting interrogation concerning the nature of elastic and diffractive 
scattering 
in strong interactions. The question is whether or not 
this 
interaction is mediated by the exchange of an object, the Pomeron,  with  properties  of 
a 
well-defined 
 hadronic particle or, at least, of a well-defined  Regge pole appearing in all 
diffractive processes.  

In this context, in a first class of models initiated
in Ref. \cite{ingelman} hard diffraction is explained 
by deep 
inelastic 
scattering (DIS) on the Pomeron, in a similar way as DIS on the proton leads to  
non-diffractive events. In a second class of models, diffractive  events are not
distinguished from non-diffractive ones, except  by a soft color interaction (SCI) 
\cite{sci} (or 
Lund string reconnection) which  may restore color singlet 
exchange. In 
this second approach, the notion of a Pomeron is a priori absent.

In the present paper we show that the forward detector apparatus 
in the D\O\ experiment at the Tevatron, Fermilab, has the 
potential to discriminate between the predictions of the two approaches in  
hard ``double'' diffractive production, e.g. of  centrally produced dijets,  by looking 
to the 
azimuthal 
distributions of the outgoing proton and antiproton with respect to the 
beam direction. 
This 
measurement relies on  
tagging  
both outgoing particles in roman pot detectors
installed by  the D0 experiment. 
We show from a Monte-Carlo simulation that this measurement 
can give significant results during  the present RUN II at the Tevatron.

\vspace{0.5cm}

{\bf 2. Theoretical framework} 
The discriminative potential of our proposal takes its origin  in the 
factorization breaking properties which were already observed at the 
Tevatron.
Both classes of models have a radically different explanation for this factorization 
breaking, cf. Fig.1.

The  Pomeron hypothesis implies the  Regge factorization property, the
same Pomeron vertex can be used to compute different diffractive
processes, e.g. the proton vertex at HERA and 
the Tevatron. In fact, hard diffraction 
at the 
Tevatron, e.g. diffractive dijet production, has revealed strong violations of 
factorization 
in hard diffraction \cite{violation}. The explanation given to this factorization 
breaking is the occurrence of large corrections from the  survival 
probabilities, which is the probability to keep a diffractive event
signed either by tagging the proton in the final state or by requiring the
existence of a rapidity gap in the event. 

The soft scattering  between incident particles tends to mask the genuine
hard diffractive interactions at 
hadronic colliders. The formulation of this correction \cite{sp} to the
scattering amplitude ${\cal A}$ consists in considering a gap 
survival  probability ($SP$) function $S$ such that 
\begin{equation}
{\cal A}(p_{T1},p_{T2}, \Delta \Phi) =
\left\{ 1 +{\cal A}_{SP} \right\}{\bf *} {\cal A}^{h}\equiv {\cal S} {\bf *} 
{\cal A}^{h} = \int d^2{\bf k}_T\ {\cal S}({\bf k}_T) \ {\cal A}^{h}({\bf p}_{T1}\!-\!{\bf k}_T,
{\bf p}_{T2}\!+
\!{\bf k}_T) 
\ ,  
\label{sp}
\end{equation}
where ${\bf p}_{T1,2}$ are the transverse momenta of the outgoing $p,\bar p$ and $\Delta \Phi$ their 
azimuthal angle separation. In our study the hard scattering 
amplitude ${\cal A}_{h}$ is obtained from
the factorizable Pomeron model POMWIG \cite{pomwig}. 
${\cal A}_{SP}$ is the soft 
scattering amplitude. In our simulations we used  two  
different models, either the two-channel eikonal {\it model~1} \cite{durham} (elastic and  
low-mass diffraction) or only the elastic channel {\it model 2} as proposed for hard diffraction in \cite{bialas}. 

By contrast with Pomeron models, 
soft color interaction models are by nature non factorizable. 
As described in Fig.1, 
the initial hard interaction is the generic standard QCD dijet production, 
accompanied by the full parton shower. Then, a phenomenological soft color 
interaction is assumed 
to modify the overall color content, allowing for a color singlet 
exchange and thus diffraction. 
This process is evaluated using a Monte-Carlo simulation 
\cite{sci_mc} which we used in our study.


\vspace{0.5cm}

{\bf 3. The D\O\ Forward Proton Detector} 

The Forward Proton Detector (FPD) \cite{fpd} installed by the D\O\
collaboration provides a unique opportunity to measure the azimuthal 
angle $\Phi$ of the outgoing protons and antiprotons and thus to test 
the dependence of diffractive events at the Tevatron on
$\Delta \Phi$ between the tagged protons and antiprotons. 

The FPD consists of eight momentum spectrometers located
close to a quadrupole magnet of the Tevatron (in short
{\it quadrupole} spectrometers)
and one spectrometer close to a
dipole magnet (in short {\it dipole} spectrometer), see Fig.2.
Four quadrupole spectrometers are located on the
outgoing proton side, the other four on the antiproton side. On each side,  
the quadrupole spectrometers are
placed both in the inner (Q-IN), and outer (Q-OUT) sides of the accelerator ring,
as well as in the upper (Q-UP) and lower (Q-DOWN) directions.
They provide almost full coverage in  $\Phi$.
The dipole spectrometer, marked as D-IN in Fig.2, is placed in the inner side of the ring in the direction of outgoing antiprotons.

Each spectrometer allows one to reconstruct the
trajectories of outgoing protons and antiprotons near the beam pipe
and thus to measure their energies and scattering angles. Spectrometers provide
high precision measurement in $t = -p_{T}^2$ and $\xi = 1 - P^\prime / E$ 
variables,
where $P^\prime$ and $p_T$ are the total and transverse momenta of the outgoing 
proton or antiproton, and $E$ is the beam energy.
The dipole detectors show a good acceptance
down to $t=0$  for $\xi > 3. 10^{-2}$ while the quadrupole detectors
are sensitive to outgoing particles down to $|t|= 0.6$ GeV$^2$ for
$\xi < 3. 10^{-2}$. This allows to obtain a good acceptance for
high mass objects diffractively produced  in the D\O\ main detector. 
For our analysis, we use a full simulation of 
the FPD acceptance in $\xi$ and $t$ \cite{fpdaccep}.

Two sorts of combinations are possible with the FPD.
In the first one, the dipole detector on the antiproton side can
be combined with a quadrupole detector on the proton side.
This combination gives {\it asymmetric} cuts on $t$ due to the
different acceptance of the two kinds of spectrometers.
The good coverage in $\Phi$ of the four quadrupole
spectrometers enables to measure the diffractive cross section 
as a function of $\Delta \Phi$ between the outgoing protons
and antiprotons. 
In the second configuration, quadrupole detectors can
be used on both sides which allows to get {\it symmetric} cuts on $t$.

\vspace{0.5cm}

{\bf 4. $\Delta \Phi$ dependence of the double diffractive
cross section} 

In Fig.3, we give the profile of the $\Delta\Phi$ dependence of the diffractive cross section. 
As an example, we require events with two jets with a 
transverse momentum greater than 5 GeV  and 
tagged proton and antiproton. The SCI model \cite{sci_mc} has been
produced using a  modified version of PYTHIA \cite{pythia}. 
The Pomeron model
has been generated using POMWIG \cite{pomwig} and the Pomeron structure
function measured by the H1 Collaboration \cite{h1pom} interfaced with the
two models for the survival probabilities described in Section 2.

We first display (upper curves) the result for asymmetric cuts in $t$ 
($|t_{p}| > 0.6$, $|t_{\bar{p}}| > 0.1$ GeV$^2$). We notice that the result
for SCI is independent on $\Delta \Phi$ whereas 
the POMWIG results with survival
probabilities show less events at high $\Delta \Phi$ by a factor of about 5.
Both survival probability models exhibit strong $\Delta \Phi$ dependence
with similar shape but with different relative normalization.
The lower plots in Fig. 3 show the results for symmetric cuts
on~$t$ ($|t_{p, \bar{p}}| > 0.5$~GeV$^2$). The difference
between SCI and POMWIG models is even larger
in this configuration, and goes up to a factor 30. Both survival probability 
models show similar behavior but the 
position of the minimum in $\Delta \Phi$ is slightly shifted.

\vspace{0.5cm}

{\bf 5. Proposed measurement at the Tevatron}

The first measurement we propose, and which can be performed even at low 
luminosity, directly benefits  from the 
FPD configuration, i.e. from the structure in $\Phi$ of the detector itself.
We suggest to count the number of events
with tagged $p$ and $\bar{p}$ for different combinations of FPD spectrometers.
For this purpose, we define the following
configurations for dipole-quadrupole tags
(see Fig. 2): same side (corresponding to D-IN on $\bar{p}$ side
and Q-IN on $p$ side and thus to $\Delta \Phi < 45$ degrees), 
opposite side (corresponding to D-IN on $\bar{p}$ side
and Q-OUT on $p$ side, and thus to $\Delta \Phi > 135$ degrees),
and middle side (corresponding to D-IN on $\bar{p}$ side
and Q-UP or Q-DOWN on $p$ side and thus to $45 < \Delta \Phi < 135$ degrees).
We define the same kinds of configurations for quadrupole-quadrupole tags
(for instance, the same side configuration corresponds to the sum of the four
possibilities: both protons and antiprotons tagged in Q-UP, Q-DOWN,
Q-IN or Q-OUT).

In Table 1, we give the ratios $1/2 \times middle/same$ and $opposite/same$ 
($middle$ is divided by 2 to get the same domain size in
$\Phi$) for the different models. In order to obtain these predictions, we used
the full acceptance in $t$ and $\xi$ of the FPD detector \cite{fpdaccep}.
Moreover we computed the ratios for two  different tagging configurations
for the symmetric and asymmetric cuts in $t$ described above,
namely for $\bar{p}$ tagged
in dipole detectors, and $p$ in quadrupoles, or for both $p$ and $\bar{p}$ 
tagged in quadrupole detectors.

In Table 1, we observe that the $\Delta \Phi$ dependence of the
event rate ratio for the SCI model 
is weak, whereas for the POMWIG models the result show important differences 
specially when both $p$ and $\bar{p}$ are tagged in quadrupole
detectors. This measurement can be performed even at low luminosity. Indeed,
the expected number of events for POMWIG for 10 pb$^{-1}$ is 
respectively about 10$^3$ (resp. about 25)
for the dipole-quadrupole (resp. quadrupole-quadrupole) configurations
if two jets with a transverse momentum greater than 5 GeV are required. This
corresponds to a very low luminosity at the Tevatron 
(about 1 week of running now), and thus
it is possible to increase the cut on the jet $p_T$ to perform this
study.

The measurement can also be performed using vector mesons ($J/ \Psi$ for
instance), or even $W$ and $Z$ at higher luminosity.   
With more luminosity, we also propose to measure directly the 
differential $\Delta \Phi$ dependence
between the outgoing protons and antiprotons using the good coverage of the
quadrupole detectors in $\Phi$ which will allow to perform a more
precise test of the models.

\vspace{0.5cm}

{\bf 6. Conclusion}

To summarize, 
we propose a new measurement to be performed at the Tevatron which can be
decisive to distinguish between Pomeron-based and soft color interaction
models of hard diffractive scattering.
The difference in azimuthal angle between the leading outgoing proton 
and antiproton in hard double diffractive interactions is found to be
a discriminating observable to distinguish between these 
two classes of models and thus to investigate the nature of the Pomeron.
We showed that this measurement can be performed with the present
D\O\ detector.

If one finds a strong $\Delta\Phi$ dependence, the soft color interaction
approach would be disfavoured unless new important changes in the way PYTHIA
deals with non-perturbative color reconnection are introduced.
On the other hand if the $\Delta\Phi$ dependence is weak, it would
mean that Pomeron concept has to be revised.

The measurement is also fundamental to obtain precise predictions
for diffractive cross section at the LHC, such as the cross section
for diffractive Higgs boson production.

\vspace{0.5cm}

{\bf Acknowledgments}

We thank Rikard Enberg for useful discussions and Jorge Barreto 
for providing the simulated acceptance of the FPD.

\begin{table}
\begin{center}
\begin{tabular}{|c|c||c|c|} \hline
 Configuration & model & middle/same & opposite/same \\ 
\hline\hline
Quad. $+$ Dipole & SCI & 1.3 & 1.1 \\
               & Pomeron Model 1 & 0.36 & 0.18 \\
               & Pomeron Model 2 & 0.47 & 0.20 \\ \hline
Quad. $+$ Quad. & SCI & 1.4 & 1.2 \\
               & Pomeron Model 1 & 0.14 & 0.31 \\
               & Pomeron Model 2 & 0.20 &  0.049     
\\
\hline
\end{tabular}
\end{center}
\caption{{\it Predictions for a proposed measurement of diffractive 
cross section ratios in different regions of $\Delta \Phi$ at the Tevatron}
(see text for the definition of middle, same and opposite).
The first (resp. second) measurement involves the dipole and 
one quadrupole detectors (resp. quadrupole detectors only) corresponding
 to asymmetric (resp. symmetric) cuts on $t$.}
\end{table}

\begin{figure}
\epsfig{file=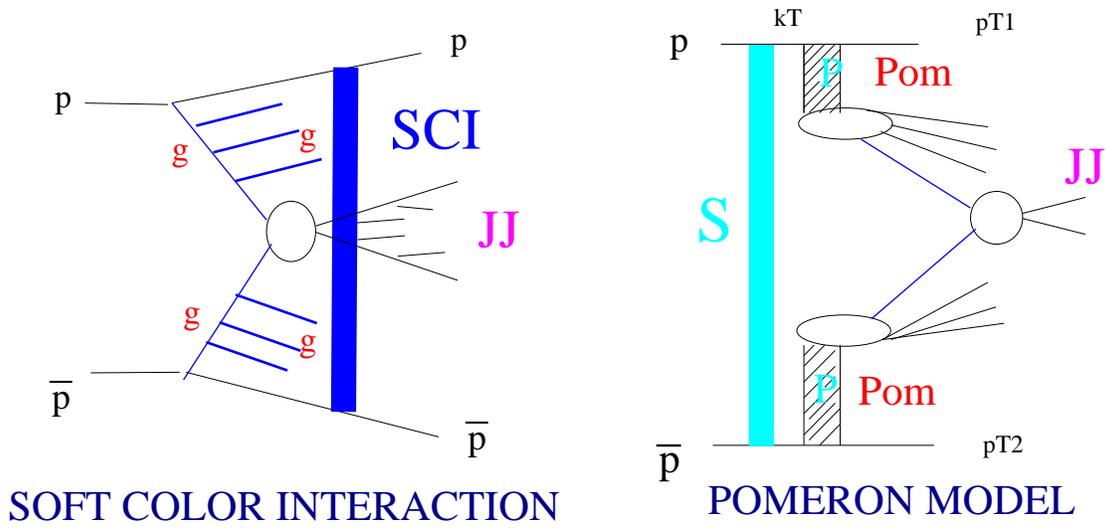,width=15cm}
\caption{{\it Description of the SCI and Pomeron models for dijet (JJ) 
diffractive production.} Left scheme: SCI model; the standard QCD dijet
production is modified by the soft color interaction (SCI). 
Right scheme: Pomeron model; the factorized double Pomeron dijet production
is corrected for the initial
soft interaction ${\cal S}$, see text.}
\label{fig1ex}
\end{figure}

\begin{figure}
\epsfig{file=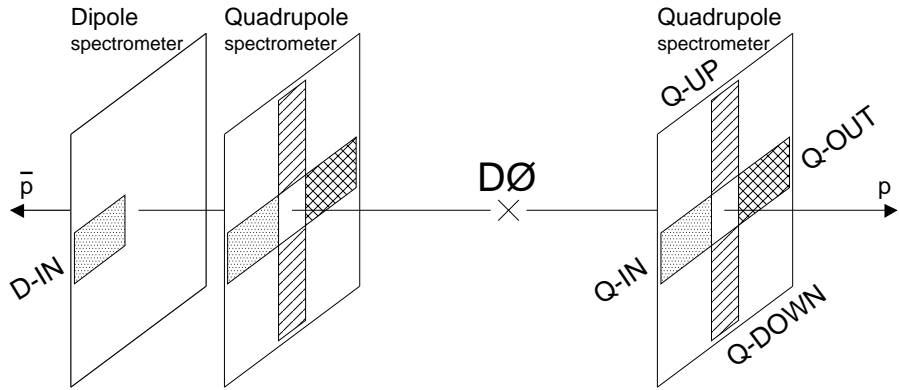,width=12cm}
\caption{{\it Schematic view of the FPD detector.} We show the positions of the
dipole and quadrupole spectrometers with respect to the main D\O\ detector.
The quadrupole detectors on $p$ and $\bar{p}$ sides consist of
4 spectrometers called Q-UP, Q-DOWN, Q-IN, Q-OUT, and the dipole detector on
$\bar{p}$ side only of one spectrometer D-IN.}
\label{fpd}
\end{figure}

\begin{figure}
\epsfig{file=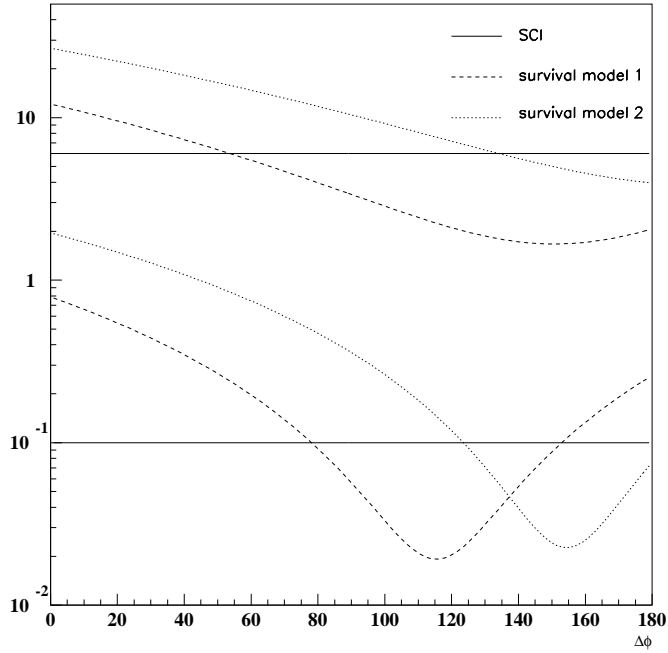,width=10cm}
\caption{{\it Predicted profile of $\Delta \Phi$ between the 
outgoing $p$ and $\bar{p}$ for SCI and Pomeron-based models.} 
The upper curves are for asymmetric cuts in $t$ 
($|t_{p}| > 0.6$, $|t_{\bar{p}}| > 0.1$ GeV$^2$) and the lower ones
for symmetric cuts
on $t$ ($|t_{p,\bar{p}}| > 0.5$~GeV$^2$). Solid lines: SCI model, 
dashed lines: Pomeron {\it model 1}, and dotted lines: Pomeron
{\it model 2} (see text).
Note that for Pomeron models the minimum is close to back-to-back
proton and antiproton for asymmetric cuts while it is around 130 degrees
for symmetric cuts.}
\label{prl2}
\end{figure}


\end{document}